\documentclass[aps,pre,twocolumn,amssymb,10pt]{revtex4-1}

\usepackage{graphicx}  
\usepackage{dcolumn}   
\usepackage{bm}        
\usepackage{amssymb}   
\usepackage{amsmath}   
\usepackage{mathtools} 
\usepackage{color}
\usepackage{ulem}
\usepackage{caption}
\usepackage{subcaption}

\usepackage[colorlinks,linkcolor=blue,urlcolor=blue,citecolor=blue]{hyperref}

\begin{document}

\def\be{\begin{equation}}
\def\ee{\end{equation}}
\def\bea{\begin{eqnarray}}
\def\eea{\end{eqnarray}}
\def\l{\label}

\newcommand{\eref}[1]{Eq.~(\ref{#1})}%
\newcommand{\Eref}[1]{Equation~(\ref{#1})}%
\newcommand{\fref}[1]{Fig.~\ref{#1}} %
\newcommand{\Fref}[1]{Figure~\ref{#1}}%
\newcommand{\sref}[1]{Sec.~\ref{#1}}%
\newcommand{\Sref}[1]{Section~\ref{#1}}%
\newcommand{\aref}[1]{Appendix~\ref{#1}}%
\newcommand{\sgn}[1]{\mathrm{sgn}({#1})}%
\newcommand{\erfc}{\mathrm{erfc}}%
\newcommand{\erf}{\mathrm{erf}}%
\newcommand{\OS}[1]{\textcolor{magenta}    {\bf O.S.: #1}}

\title{
Thermodynamic uncertainty relations for many-body systems with fast jump rates and large occupancies
}
\author{Ohad Shpielberg$^{1}$}
\email{ohads@sci.haifa.ac.il}
\affiliation{$^{1}$Haifa Research Center for Theoretical Physics and Astrophysics, University of Haifa, Mt. Carmel, Haifa 31905, Israel}

\author{Arnab Pal$^{2}$}
\email{arnabp@iitk.ac.in}
\affiliation{$^{2}$Department of Physics, Indian Institute of Technology, Kanpur, Kanpur 208016, India}

\date{\today}

\begin{abstract}
\noindent
A universal large $\mathcal{N}$ theory of nonequilibrium fluctuations emerges in the limit of fast jump rates and large occupancies. We use this theory to derive a set of coarse grained thermodynamic uncertainty relations (TUR) -- one of them being an activity bound.  Importantly, the activity serves as a tighter bound for the entropy production in 1D systems. These results are particularly useful in the many-body regime, where typically a coarse grained approach is required to handle the large microscopic state space.
\end{abstract}

\maketitle

\section{Introduction}

The second law of thermodynamics resulting in, e.g. the Carnot bound on the maximum efficiency of thermal engines, demonstrates the importance that inequalities play in physics. The Carnot efficiency bound is remarkably independent of specific design. More recently and in the same spirit, the thermodynamic uncertainty relations (TUR) \cite{Barato_Seifert15,Gingrich_PRL16,horowitz2020thermodynamic} revealed that fluctuations in thermal systems cannot be freely minimised.  Rather they are bounded from below by the inverse entropy production irrespective of system design.  The ideas of the TUR led to an effort towards optimizing the bounds \cite{UnidrectionalTUR,UnifiedTUR_Kashia,Shiraishi_Optimalbound,pal2021thermodynamic},  generalizing the bound in the regime of large deviations \cite{Pietzonka_LDF,TUR_FPT},  quantum systems \cite{Hasegawa_activity_open_quantum,TUR_quantumHeatEngine,Hasegawa_ContMsrmnt,Segal_QTUR}, explicit time-dependence \cite{FiniteTime_TUR}, athermal analogues \cite{Soret20} and results of the same spirit \cite{Shiraishi_SpeedLimits,UnifiedTUR_Falasco,TUR_cycleCurrents,ImitatingNoneq_inequality,manikandan2018exact}.

A common starting point for discussing the TUR is via a master equation. The master matrix of the rates may be time-dependent or not. This language is particularly suited for a single particle dynamics, where a state corresponds to the particle localised in a given site. 
For many body systems, the applicability of the TUR is limited due to the large state space. Namely, evaluating the particle densities involves finding the zero eigenvalue state of a large Markov matrix. Similarly, evaluating current fluctuations requires finding the largest eigenvalue of a large tilted Markov matrix \cite{DerridaDoucot} (also see detailed discussion later). 
This renders the overall procedure tedious and quite often intractable. 

Indeed, it is only in $1D$ non-equilibrium steady state that the current and current fluctuations are evaluated along a single bond only. But for any generic network, accounting for a large number of states, calculating, measuring or obtaining numerically the fluctuations of the current is typically hard: thus limiting the usefulness of the TUR and its large deviation bound counterpart. This difficulty can be overcome using universal non-equilibrium theories that result in a coarse grained picture, allowing a compact description to calculate current statistics \cite{MFT_Review,Maes_Netocny,mFT_Monthus,Spohn_NFH,Meerson_geoOptics}. Naturally, the question arises on whether one can write a useful TUR in a coarse grained manner. 

In this work, we show how this can be done in a systematic way from a nonequilibrium theory: Consider a master equation with fast rates and large particle occupancies on a finite graph  \cite{Gabrielli_LargeN,Baek_largeN}. The fast-rates-large-occupancy large $\mathcal{N}$ limit leads to a universal coarse grained non-equilibrium theory, dubbed here -- the large $\mathcal{N}$ theory (see also \cite{mFT_Monthus}). Within the framework of this large $\mathcal{N}$ theory, we show that the variance of the current is bounded from below by either the activities or the coarse grained entropy production. Interestingly, the activity serves as an upper bound for the entropy production and a tighter bound in $1D$ systems. The latter bound becomes tight in the large $\mathcal{N}$ limit hence serving as a better tool to infer entropy production.  These results, reinstate the importance and relevance of the TUR and similar inequalities in many body systems even in the case where the states space is unmanageable to treat, analytically or numerically.

The outline of this work is as follows. In \sref{sec:Setup}, we lay the setup for the large $\mathcal{N}$ theory and present the main results. In \sref{sec:ASIP}, we numerically validate the main results for a particular interacting model system: the asymmetric inclusion process (ASIP) at finite $\mathcal{N}$ \cite{Reuveni_ASIP,Grosskinsky_ASIP}. \sref{sec:Derivations} focuses on the derivation of the bound and the results of \sref{sec:Setup}. We conclude in \sref{Sec:Discussion}
 where we summarise the work and point out future directions.


\section{Setup and results}
\label{sec:Setup}



Consider a stochastic process with
a finite set of sites with particle occupancies $n_x =0,1,2,...$.
Assume that a particle jumps from site $x$ to $y$ with rate $\tilde {\kappa}_{y,x}$ that may depend on the local occupancies.  Particles are added to / removed from the system only through a finite number of bonds to a reservoir or a set of reservoirs. 
In particular, we restrict the discussion to processes with fast rates; scaling like  $\mathcal{N}^2$ with respect to the large parameter $\mathcal{N}\gg 1$. Under this assumption, we consider the rescaled occupancies $\rho_x (\tau) = n_x /\mathcal{N} $ and the rescaled time $\tau = \mathcal{N}t$. Then,  $\tilde{\kappa}_{y,x} = \mathcal{N}^2 \kappa_{y,x} + O(\mathcal{N})$, where $\kappa_{y,x}$ depends only on the rescaled densities $\rho$. Gathering these definitions, the rescaled evolution equation is
\begin{eqnarray}
    \partial_{\tau} \rho_x &=& - (\mathrm{div } \, \kappa )_x ,
    \\ \nonumber 
    (\mathrm{div } \, \kappa   )_x  
      &=& - \sum_{y\sim x } \kappa_{x,y} -  \kappa_{y,x},
\end{eqnarray}
where $\sum_{y \sim x} $ denotes summation over all the neighbors of $x$. Note that this summation may include the coupling to reservoirs which are assumed to have fixed densities. We now define the empirical unidirectional flux  $\tilde{q}_{x,y}(t)$  that counts the number of particles jumping from site $y$ to $x$ during the time interval $\left[ t,t+dt\right]$. At the large $\mathcal{N}$ limit, the rescaled empirical unidirectional flux is  $\tilde{q}_{y,x}(t) = \mathcal{N}^2q_{y,x}(\tau) +O(\mathcal{N})$. Note that, on average, $\langle q_{x,y}(\tau) \rangle = \langle \kappa_{x,y}(\rho(\tau)) \rangle = \kappa_{x,y}( \langle \rho(\tau)\rangle   )$. The last equality uses the mean field approximation, which will be shown to be exact at $\mathcal{N}\rightarrow \infty$. The notation $\kappa(\rho)$ implies that $\kappa$ depends on the local densities. The (rescaled) empirical current is given by $j_{x,y}(\tau)  \equiv q_{x,y}(\tau ) - q_{y,x}(\tau) $. We further define $J_{y,x} (\rho) \equiv \kappa_{y,x} - \kappa_{x,y}$. Thus, on average, $ \langle j_{y,x} \rangle = \langle J_{y,x}( \rho )\rangle = J_{y,x}(\langle \rho \rangle)  $ and again the last inequality is exact only in the limit $\mathcal{N}\rightarrow \infty$. Lastly, we define the local activity $a_{x,y} \equiv q_{x,y}+ q_{y,x}$ \cite{Maes_Netocny} that will play an important role later on.

In what follows, we explore the properties of a generalized fluctuating charge transfer defined in the following
\begin{equation}
 Q = \mathcal{N} \int^\tau _0 d\tau' \, \sum_x \sum_{y>x} d_{y,x} j_{y,x} + \sum_x u_x \rho_x, \end{equation}
 where $d_{y,x}, u_x$ are predetermined weights. We furthermore define $d_{x,y} = - d_{y,x}$ for convenience. It has no bearing on $Q$. Note that the functional contains both the current and occupation like terms. 
 In this work, we derive a bound for the cumulant generating function $  \mu(\lambda)     \equiv   \frac{1}{\mathcal{N}\tau} \ln \langle e^{\lambda Q} \rangle $ of the charge $Q$ at the steady state. In particular, the bound on the generalized current variance reads
 \begin{eqnarray}
 \label{eq:m2 activity bound}
 \mu''(0) &\geq& (d^\perp)^2  
    \langle
    A^\parallel  \rangle  ,
\end{eqnarray}
where $(A^\parallel)^{-1} \equiv  \sum_{x,y>x} \frac{1}{ a_{y,x} }$ -- an impedance-like parallel summation over the activities, and $d^\perp \equiv \sum_{x,y>x} d_{y,x}$ -- an impedance-like series summation over the weights. Detailed derivation of our results is presented in Sec. \ref{sec:Derivations}. In the large $\mathcal{N}$ theory, $\langle a^{-1} _{x,y} \rangle = \langle a _{x,y} \rangle^{-1} $ and the mean bond activities depends only on the mean densities. The lower bound is accessible analytically and numerically. Moreover, even if the rates are not known, the activities -- a measure on the number of jumps -- are still accessible experimentally in many cases \cite{Visual_Activity,svetlizky2021spatial}. 
Interestingly, the activity bound \eqref{eq:m2 activity bound} can further be generalized to a large deviation bound \eqref{eq: Y activity LDF bound} that we show later.

In particular for a 1D lattice, we can improve the well known entropy production bound by using the parallel activity
\begin{eqnarray}
\label{eq:series of bounds}
\mu''(0) &\geq& (d^\perp)^2 \langle A^\parallel \rangle \geq 
    \frac{2 (\mu' _j )^2}{\langle \Sigma \rangle   } ,
\end{eqnarray}
with $\mu' _j \equiv  \sum_{x,y>x} d_{y,x}\langle j_{y,x}\rangle $ and
\begin{align}
 \Sigma = \sum_{x,y>x}  j_{y,x} \log \frac{a_{y,x}+j_{y,x}}{a_{y,x}-j_{y,x}} ,   
\end{align}
as the entropy production rate \cite{seifert2012stochastic,van2015ensemble}.
A similar series of bounds as in \eqref{eq:series of bounds} was already derived in the case of a 1D periodic system governed by a master equation \cite{Barato_Gabrielli}. Notice that \cite{Barato_Gabrielli} was evaluated through the microscopic master equation itself whereas in our case, we are interested in the coarse grained quantities, e.g. the densities $\rho$ and the rescaled rates $\kappa$.  Therefore, the coarse grained series of inequalities need not be exact at finite $\mathcal{N}$ and deviations from it are observed, but controlled. See Fig.\ref{fig:mu A difference}. As noted before, \eqref{eq:series of bounds} suggests the bound is given in terms of the mean local activities which are accessible numerically, analytically and usually also experimentally when dealing with a finite graph.

Lastly, let us use \eqref{eq:series of bounds} to imply two more appealing bounds. Notice that $A^\parallel \leq  A^\perp / (\sum_{x,y>x} 1 )^2 $ by Titu's lemma \cite{sedrakyan2018algebraic} where $A^\perp$ is the activity series summation. Furthermore,  $A^\parallel \leq a_{x,y}$ for any bond pair $x,y$. This leads to two bounds 
\begin{align}
\label{eq:two useful EP bounds}
    \langle \Sigma \rangle &\geq \frac{2 (\mu'_j)^2 (\sum_{x,y>x} 1)^2}{(d^\perp)^2 \langle A^\perp \rangle },
    \\ 
    \langle \Sigma \rangle &\geq \frac{2 (\mu'_j)^2 }{(d^\perp)^2 \langle a_{x,y} \rangle } \quad \forall x,y.
\end{align}
The first bound limits the entropy production in terms of the activity. This bound does not bear the same content as the kinetic uncertainty relation \cite{KUR_Baiesi} as it bounds the entropy production and \textit{not} the current variance. The second inequality is particularly interesting for practical purposes.  It allows to get a lower bound on the entropy production from a single bond current and activity. In 1D systems, the steady state current is uniform for any bond i.e. $\mu' _j = d^\perp \langle j_{y,x} \rangle, \forall \, y>x$.  The activity, however, is not uniform which makes this result surprising.

 \section{Application: the ASIP}
 \label{sec:ASIP}

\begin{figure}
     \centering
     \begin{subfigure}[b]{0.3\textwidth}
         \centering
         \includegraphics[width=\textwidth]{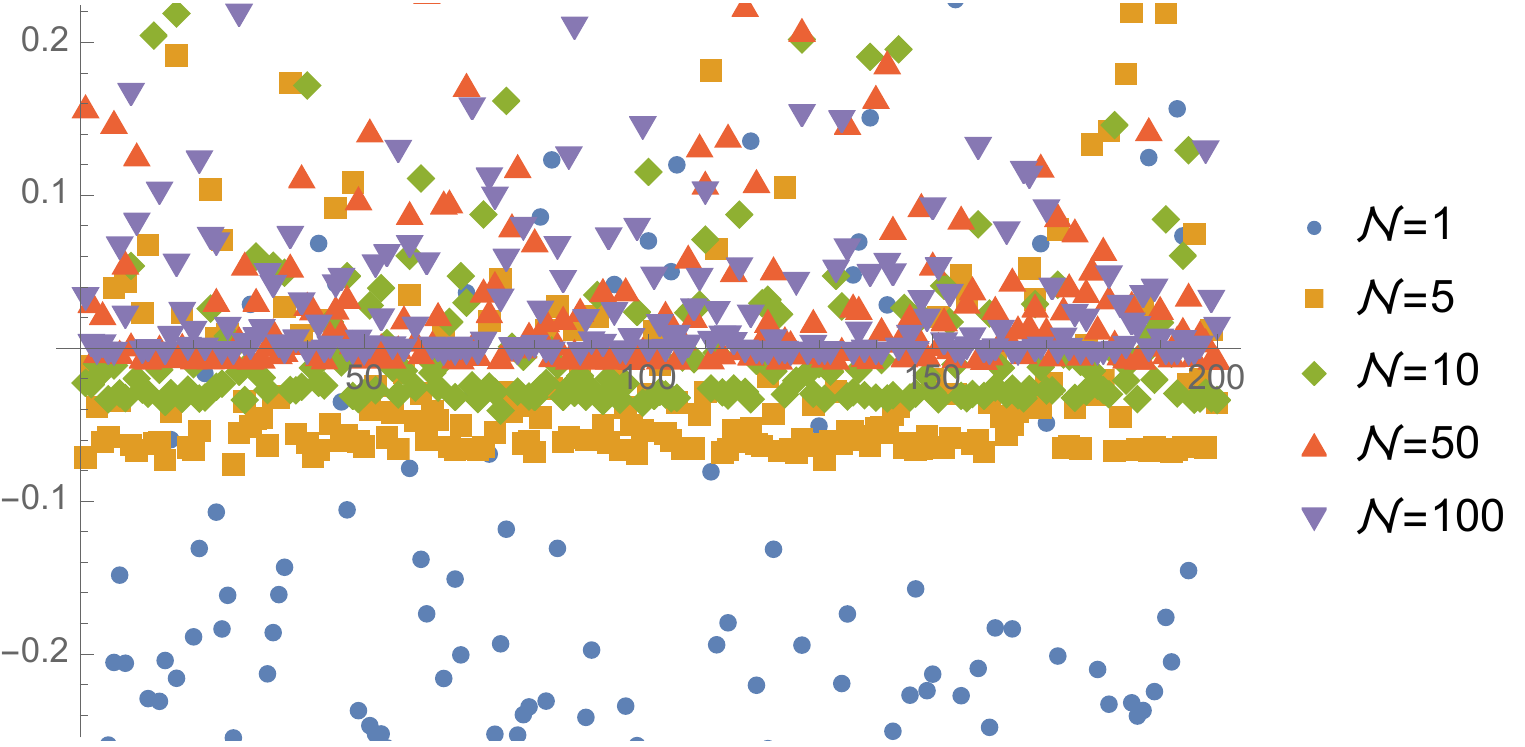}
         \caption{$200$ realizations of the ASIP with random rates $p_\pm(x) \in \left[ 0,1 \right]  $ where $N=2$ particles and $L=3$ sites with periodic boundary conditions. }
     \end{subfigure}
     \hfill
     \begin{subfigure}[b]{0.3\textwidth}
         \centering
     \includegraphics[width=\textwidth]{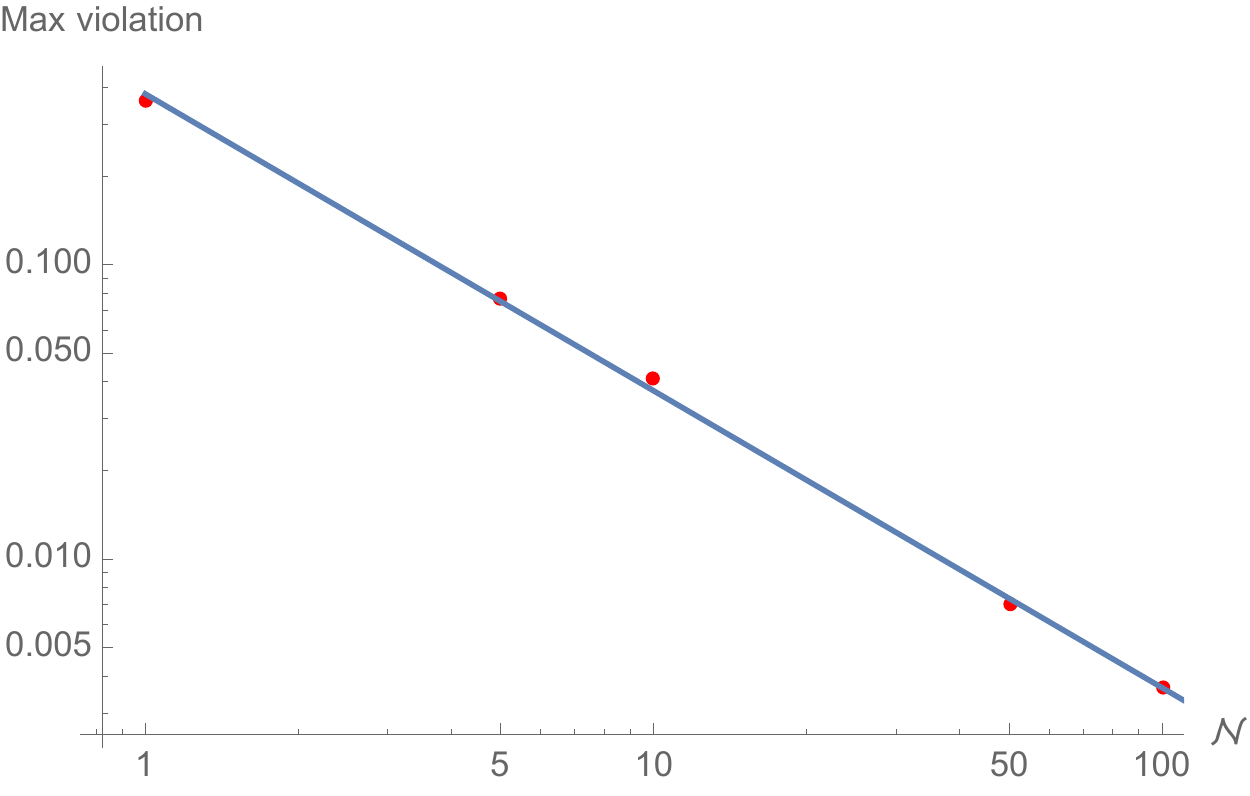}
         \caption{The negative of minimal values of $\frac{\mu''(0)-(d^\perp)^2 \langle A^\parallel \rangle }{\mu''(0)}$ out of all the random realizations for each $\mathcal{N}$ -- the max violation.  }
     \end{subfigure}
     \hfill
        \caption{Numerical demonstration of the first inequality in \eqref{eq:series of bounds}. The value of $\frac{\mu''(0)-(d^\perp)^2 \langle A^\parallel \rangle }{\mu''(0)}$ in the random ASIP. We present $200$ random realizations for each $\mathcal{N}$ value.  
        (a) The minimal values of the random realizations are seen to converge to zero with increasing $\mathcal{N}$. (b) The violation -- the negative of the minimal value of all the realizations -- is shown to vanish like $1/\mathcal{N}$, as expected from the large $\mathcal{N}$ theory. The fitting gives a slope of $-1.008$ close to the expected slope of $-1$ at $\mathcal{N}=\infty$. For more details, see Sec. \ref{app:Numerically exploring the bounds}.}
        \label{fig:mu A difference}
\end{figure}

\begin{figure}
     \centering
     \begin{subfigure}[b]{0.3\textwidth}
         \centering
         \includegraphics[width=\textwidth]{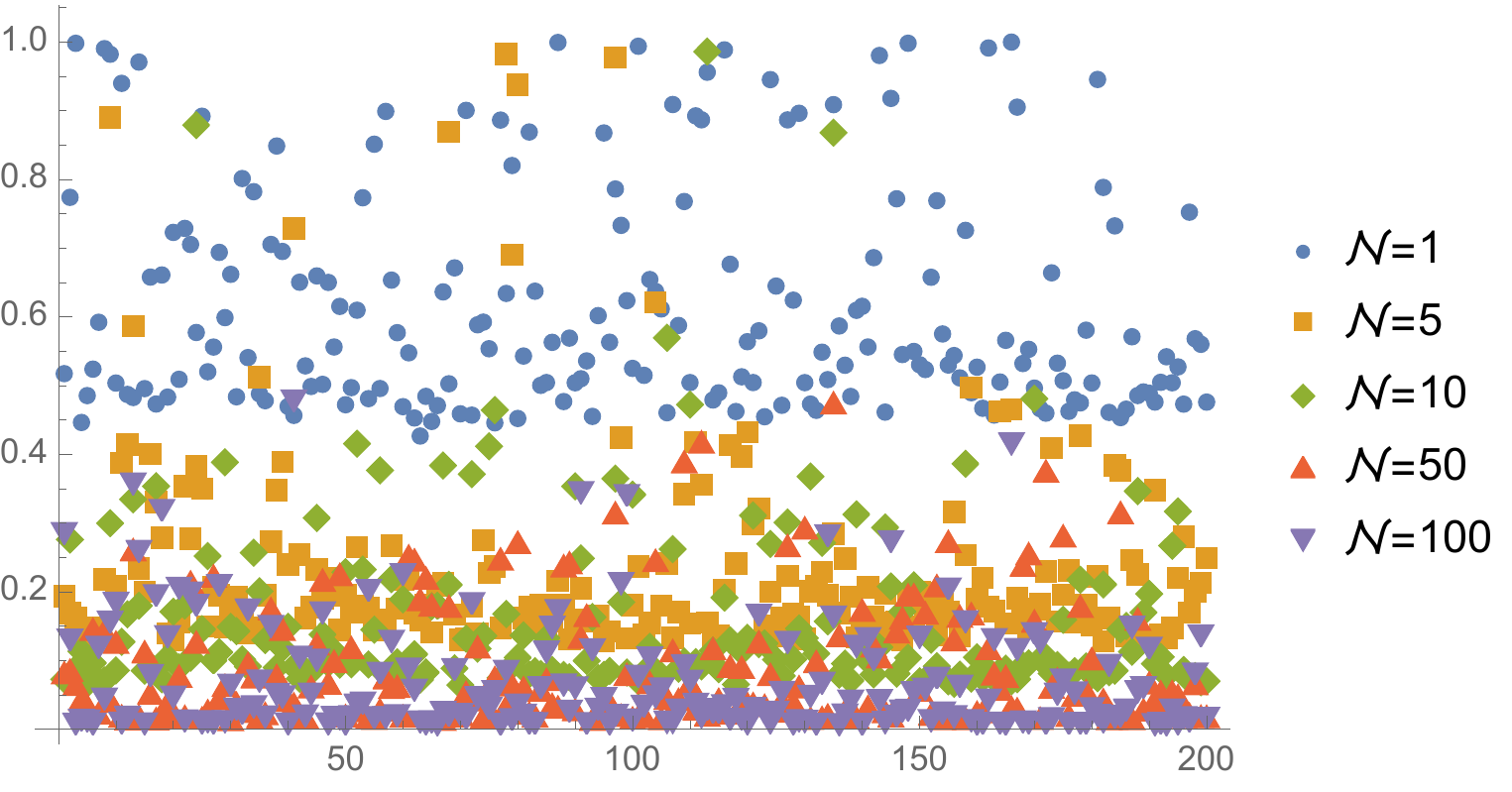}
         \caption{ $200$ realizations of the ASIP with random rates $p_\pm(x) \in \left[ 0,1 \right]  $ where  $N=2$ particles and $L=3$ sites with periodic boundary conditions.}
     \end{subfigure}
     \hfill
     \begin{subfigure}[b]{0.3\textwidth}
         \centering
         \includegraphics[width=\textwidth]{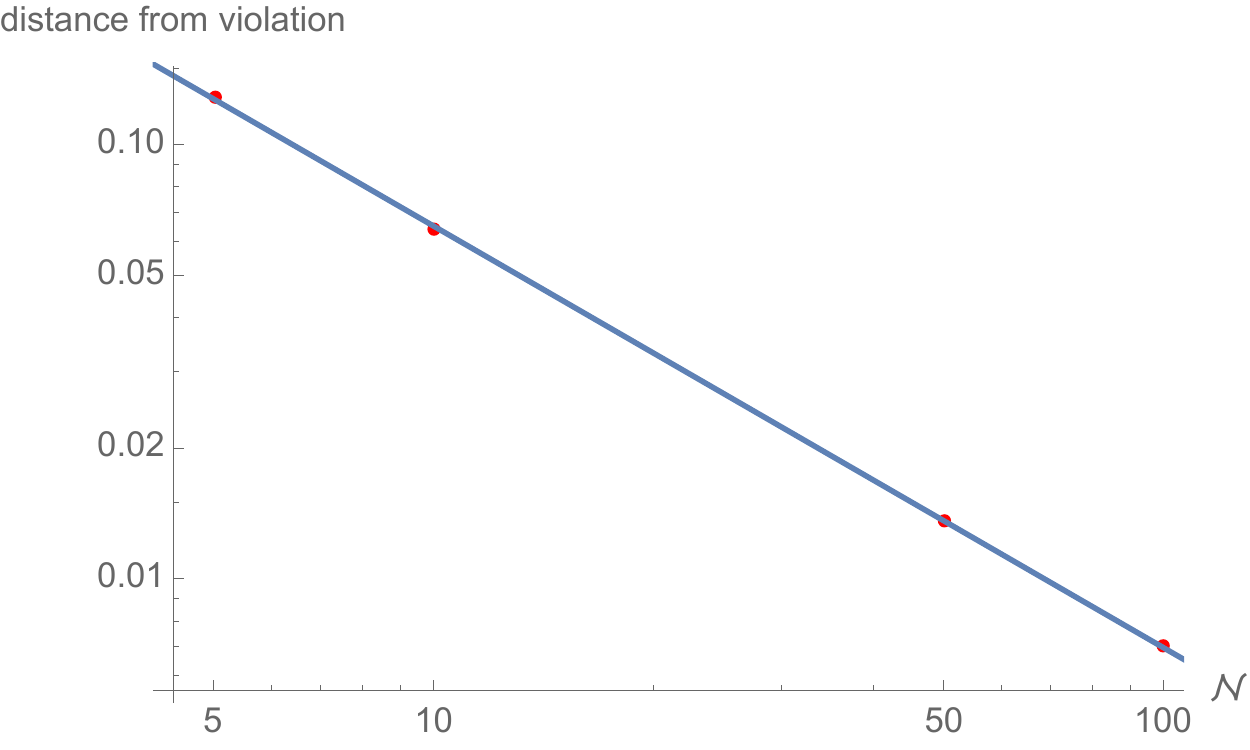}
         \caption{ The minimal value of $\frac{(d^\perp)^2 \langle A^\parallel \rangle - 2 (\mu' _j)^2/\langle \Sigma \rangle  }{(d^\perp)^2 \langle A^\parallel \rangle}$ out of all the random realizations for each $\mathcal{N}$. Here, this is the distance from violation of the bound.   }
     \end{subfigure}
     \hfill
        \caption{Numerical demonstration of the second inequality in \eqref{eq:series of bounds}. The evidence of $\frac{(d^\perp)^2 \langle A^\parallel \rangle - 2 (\mu' _j)^2/\langle \Sigma \rangle  }{(d^\perp)^2 \langle A^\parallel \rangle}\geq 0$ in the random ASIP.  We present $200$ random realizations for each $\mathcal{N}$ value.  
        (a) The minimal values of the random realizations are seen to converge to zero with increasing $\mathcal{N}$ indicating the tightening of the bound. (b) The distance from violation -- the minimal value of all the realizations -- is shown to vanish like $1/\mathcal{N}$. The fitting gives a slope of $-0.9715$ close to the expected slope of $-1$ at $\mathcal{N}=\infty$. For more details, see Sec. \ref{app:Numerically exploring the bounds}.}
        \label{fig: A sigma difference}
\end{figure}

In this section, we illustrate our results using an interacting particle system namely the Asymmetric Inclusion Process (ASIP) \cite{Reuveni_ASIP,Grosskinsky_ASIP}.
 In particular, we focus on the dynamics on a 1D chain \cite{Reuveni_ASIP,Grosskinsky_ASIP} where we have $\tilde{\kappa}_{x\pm 1,x} = p_\pm(x) n_x (\mathcal{N}+n_{x\pm1})  $ with 
 the rates
 $p_{\pm}(x)\geq 0$. Then, $\kappa_{x \pm 1, x} = p_\pm(x) \rho_x (1+ \rho_{x+1}) $.   
To validate our claims, in particular \eqref{eq:series of bounds}, we consider the ASIP with $N=2$ particles and a periodic system of $L=3$ sites and random rates $p_{\pm}(x)\in [0,1]$. We evaluate the local densities at the steady state and recover local coarse grained activities and currents.  

In Fig.\ref{fig:mu A difference}, the inequality $\mu''(0) - (d^\perp)^2 \langle A^\parallel \rangle \geq 0$ is demonstrated in the large $\mathcal{N}$ limit, with $1/\mathcal{N}$ corrections, as expected from the large $\mathcal{N}$ theory. Namely, the inequality is precise only at $\mathcal{N}\rightarrow \infty$. In Fig.~\ref{fig: A sigma difference}, the inequality $(d^\perp)^2 \langle A^\parallel \rangle  \geq 2(\mu' _j)^2 / \langle \Sigma \rangle   $ is shown to hold for any $\mathcal{N}$. This need not be generally true, and probably results from the particular choice of the ASIP dynamics. Nevertheless, this inequality is shown to become the tightest in the large $\mathcal{N}$ limit with $1/\mathcal{N}$ convergence. 
Further details on the numerical results are discussed in the Appendices \ref{app:Numerically exploring the bounds} and \ref{app: the F numerics}. 

 \section{Derivation of the bounds}
 \label{sec:Derivations}
 In this section, we derive the central results that were highlighted in Sec. \ref{sec:Setup}. We start by noting that the joint path probability for the current and density at the large $\mathcal{N}$ limit is given by $P(j,\rho) \sim \exp{(- \mathcal{N} \int d\tau \mathcal{L})} $ \cite{mFT_Monthus} (also, see the appendix \ref{app:probability distribution} and the discussion \footnote{As usual, we discard here logarithmic corrections, normalization and $1/\mathcal{N}$ corrections to the action. }).
 The Lagrangian $\mathcal{L}$ defines the path probability at the large $\mathcal{N}$ limit up to $1/\mathcal{N}$ corrections
 with 
 \begin{eqnarray}
     \mathcal{L} =  \sum_{x, y>x} \Phi(j_{y,x},\kappa_{y,x},\kappa_{x,y}),
     \label{Lagrangian}
 \end{eqnarray}
where
\begin{eqnarray}
\label{eq:2point LDF}
\Phi \left( j, \kappa_+ , \kappa_- \right)  &=& j \ln{
\frac{j+\sqrt{j ^2 + 4 \kappa_+ \kappa_-}}{2\kappa_+}
} \\ \nonumber 
&& - \sqrt{j ^2 + 4 \kappa_+ \kappa_-} + \kappa_+ +\kappa_-.
\end{eqnarray}
Note that the summation in (\ref{Lagrangian}) avoids double counting of the bonds.  The expression \eqref{eq:2point LDF} is not new, and can be found in \cite{mFT_Monthus,Gabrielli_LargeN,Baek_largeN} and of course earlier in the non-interacting limits where $\kappa_{y,x} \propto \rho_x $ \cite{Maes_Netocny}.  Note that $\mathcal{L}$ is not limited to closed systems. In open systems, the incoming/outgoing rates $\kappa_\pm$ may depend on the fixed density of the reservoirs \cite{Baek_largeN,mFT_Monthus}. In this case, the state space of the Markov matrix is unbounded and the large $\mathcal{N}$ theory is particularly useful.



It is now useful to present the Lagrangian $\mathcal{L}_Y $, which depends on the tilting field $Y_{y,x}$, constrained to satisfy Kirchhoff's junction rule  \cite{SasaDechant18,Soret20}
\begin{align}
\label{eq:Kirchhoffs rule}
\sum_{y\sim x} Y_{y,x} =0  ~~\text{and}~~  Y_{y,x} = - Y_{x,y}  .
\end{align}
The expression is defined locally $\Phi_Y(j,\kappa_+ ,\kappa_-) \equiv \Phi(j, \kappa_+ +   \frac{1}{2} Y, \kappa_- -    \frac{1}{2} Y)$ which leads to the steady state densities $\langle \rho_x \rangle = \langle \rho_x \rangle_Y$, but a different steady state current $\langle j_{y,x} \rangle_Y = \langle j_{y,x} \rangle + Y_{y,x}$    \footnote{There is some freedom in this definition as we wish to control only the steady densities and currents.}. The averaging $\langle \, \cdot \,  \rangle_Y$ is with respect to the tilted Lagrangian $\mathcal{L}_Y$. One natural choice for the tilted dynamics is to induce a mapping to equilibrium, i.e. setting $\langle j_{x,y} \rangle_Y =0 $, implying the choice $Y_{y,x}  = - \langle j_{y,x} \rangle = \langle \kappa_{x,y} - \kappa_{y,x} \rangle  $. Another natural choice is to set $Y$ such that $\langle j_{x,y}\rangle _Y = -\langle j_{x,y}\rangle $, realizing the steady state time-reversed dynamics. This in turn allows one to evaluate the entropy production --  a central quantity in the study of non-equilibrium physics.  Recently, Dechant and Sasa also used a similar idea to extend the time reversal mapping in order to possibly tighten the TUR \cite{Dechant_cont_TimeReversal}.   

Here, the tilted dynamics allows more freedom, but in turn may lose an amenable physical interpretation as $\Phi $ is defined only for non-negative $\kappa_+$. 
 Nevertheless, this mathematical trick leads to an optimized bound with a clear physical interpretation. Moreover, it allows us to  optimizes the bound on the variance of $Q$ via a set of linear equations (see \eqref{eq:nu linear set}).

The cumulant generating function can now be expressed using the tilted dynamics
\begin{equation}
\mu(\lambda ) = \frac{1}{\mathcal{N}\tau }\log \langle e^{\lambda Q + \mathcal{N}\int d\tau \,  \mathcal{L}_Y - \mathcal{L}  } \rangle_Y . \end{equation} 
Using the Jensen inequality in above leads to 
\begin{equation}
\label{eq:Jensen inequality}
    \mu(\lambda) \geq  \frac{\lambda }{\mathcal{N} \tau} \langle Q \rangle_Y  + \langle \mathcal{L}_Y-\mathcal{L}_0 \rangle_Y.
\end{equation}
Note that at this point, optimization of \eqref{eq:Jensen inequality} with respect to the tilting field $Y$ leads to a set of non-linear equations which are hard to solve in general. To produce a useful inequality on the current fluctuations, we take \eqref{eq:Jensen inequality}  together with the rescaling $Y \rightarrow \lambda Y$ \cite{Dechant_cont_TimeReversal} and expand both sides to second order in $\lambda$
\begin{equation}
\label{eq:nonphysical bound}
    \frac{1}{2}\mu''(0) \geq  \sum_{x,y>x} d_{y,x} Y_{y,x} - Y^2 _{y,x} \frac{1}{2 \langle a_{y,x} \rangle}.
\end{equation}
Recall that $ a_{y,x} = \kappa_{x,y}(\rho) + \kappa_{y,x}(\rho)  $ is the bond activity.  Eq.\eqref{eq:nonphysical bound} together with Kirchhoff's junction rule \eqref{eq:Kirchhoffs rule} compose one of the central results of this work. We stress that the explicit result could be obtained due to the the saddle point approximation in the path probability resulting from the large $\mathcal{N}$ value. For finite $\mathcal{N}$, \eqref{eq:nonphysical bound} and what follows from it, may be erroneous as demonstrated for finite $\mathcal{N}$ in Fig.~\ref{fig:mu A difference}. Taken together with the Kirchhoff's junction rule \eqref{eq:Kirchhoffs rule} -- which is a constraint, and simultaneous minimization of $Y_{x,y}$ leads to tightening the TUR.

Eq.\eqref{eq:nonphysical bound} implies that many different bounds could be obtained. In what follows we discuss two particular choices for $Y$ leading to two different bounds: the activity bound and the entropy production bound. Then we discuss how to optimize the bound using the tilting field. We conclude this section by showing that for 1D systems, the bounds can be ordered, making them particularly useful. For completeness, we also connect our derivation to the kinetic uncertainty bound \cite{KUR_Baiesi,Shiraishi_Optimalbound,UnidrectionalTUR} in the Appendix \ref{app:series activity bound}.

 \subsection{ Activity bound}
Let us first explore the simplest tilting field $Y_{y,x}  = Y$ for $y>x$ together with $Y_{y,x} = - Y_{x,y}$ satisfying Kirchhoff's rule  \eqref{eq:Kirchhoffs rule}. The resulting bound is still valid for any $Y$. Optimization of the constant $Y$ (see the right hand side of Eq.\eqref{eq:nonphysical bound}) leads to $Y= d^\perp \langle A^\parallel \rangle $ and to the activity bound \eqref{eq:m2 activity bound}. Furthermore, we can directly use this (suboptimal) choice of $Y$  in \eqref{eq:Jensen inequality}, leading to the large deviation bound 
\begin{equation}
 \label{eq: Y activity LDF bound}
    \mu(\lambda) - \lambda \mu'(0)   \geq (d^\perp)^2   \lambda^2
    \langle
    A^\parallel  \rangle - \langle \mathcal{L}_0 \rangle_{\lambda d^\perp \langle A^\parallel \rangle }. 
\end{equation}
While the right hand side of  \eqref{eq: Y activity LDF bound} seems cumbersome, it can be evaluated in a straight-forward manner using the mean local activities and currents only or through the densities if the rates $\kappa$ are known.

\subsection{Optimizing the bound}

Next, we sketch the optimization of the bound with respect to the tilting field. Define $F_Y$ such that  $\mu''(0) \geq  2 F_Y$ according to \eqref{eq:nonphysical bound} and Kirchhoff's junction rule \eqref{eq:Kirchhoffs rule}. Then, we aim to find the  maximum of 
\begin{equation}
    \label{eq:FY to opt}
    F_Y =  \sum_{x, y>x} \left( d_{y,x} Y_{y,x} - \frac{Y^2 _{y,x}}{2 \langle a_{y,x}\rangle} \right) + \nu_x \sum_{y \sim x} Y_{y,x}
\end{equation}
where $\nu_x$ are Lagrange multipliers accounting for Kirchhoff's junction rule \eqref{eq:Kirchhoffs rule}. The solution to this optimization problem is 
\begin{align}
2 F_{Y,\mathrm{max}} &= \sum_{x,y>x} d^2 _{y,x} \langle a_{y,x} \rangle  - \langle a_{y,x} \rangle^2 (\nu_x -\nu_y)^2    , \nonumber
\\ 
\label{eq:nu linear set}
  0 &= \sum_{y \sim x } (d_{y,x}+ \nu_x - \nu_y) \langle a_{y,x} \rangle  .
\end{align}
Note that one needs to first solve the set of linear equations \eqref{eq:nu linear set} to obtain $F_{Y,\text{max}}$. The size of the set is essentially the number of bonds in the graph while coupling to reservoirs enlarges this number. A different optimization scheme was carried out recently resulting in the same optimized bound  \cite{Shiraishi_Optimalbound}. Other useful bounds can be obtained from \eqref{eq:nonphysical bound}. In what follows, we present a few physically relevant bounds and explore their relation and relevance.

\subsection{ Relations between bounds}

In this subsection, we benchmark the TURs which were derived in \cite{Barato_Gabrielli,Gingrich_PRL16}, using the tilted $Y$ approach. Consider $Y_{y,x} = \alpha \langle j_{y,x} \rangle  $. This choice satisfies the Kirchhoff's rule for any constant $\alpha$ as the steady state current is divergence free. Furthermore, one can show that $2\frac{\langle j_{y,x}\rangle^2  }{\langle a_{y,x} \rangle } \leq j_{y,x} \log \frac{\langle a_{x,y}\rangle + \langle j_{x,y}\rangle }{\langle a_{x,y}\rangle- \langle j_{x,y}\rangle}  $ as $2x^2 \leq x \log \frac{1+x}{1-x}$ for $x= \frac{j}{a} \in \left[-1,1 \right]$. Simple additivity of terms imply then  $\sum_{y>x} 2\frac{\langle j_{x,y}\rangle^2}{\langle a_{x,y}\rangle} \leq \langle \Sigma \rangle $. Hence we recover $\mu''(0) \geq \alpha \mu' _j - \frac{\alpha^2}{4} \langle \Sigma \rangle $. Finding the optimal $\alpha$ leads to the entropy production bound
\begin{align}
    \label{eq:LDF bound EP}
    \mu(\lambda) - \lambda \mu'(0) &\geq \frac{\lambda^2}{2} \frac{2 (\mu' _j )^2 }{\langle \Sigma \rangle },
    \\ 
    \mu''(0) &\geq  \frac{2 (\mu' _j )^2 }{\langle \Sigma \rangle }.
\end{align}
The entropy production bound as well as the activity bound can be derived by a certain choice of the tilting field $Y_{x,y}$. Therefore, it is not only the current variance that can serve as an upper bound. Let us define $\mathcal{F}  \equiv \sum_{x,y>x} d^2 _{y,x} \langle a_{y,x} \rangle $, namely it optimizes the tilting field $Y_{x,y}$ without necessarily satisfying Kirchhoff's rule \eqref{eq:Kirchhoffs rule}. Therefore, we find that \begin{equation}
\label{eq: mathcalF inequality}
    \mathcal{F} \geq   2\frac{(\mu' _j)^2}{\langle \Sigma \rangle } , (d^\perp)^2 \langle A^\parallel \rangle  .
\end{equation} 
Eq.\eqref{eq: mathcalF inequality} could be made more physically relevant.  The non-negativity of $d^2_{y,x} ,\langle a_{y,x} \rangle $ implies   $\mathcal{F} \leq \sum_{y>x} d^2 _{y,x}  \langle A^\perp \rangle  $. Therefore, we recover from \eqref{eq: mathcalF inequality} the physical bound
\begin{equation}
\label{eq: gen series activity bound}
    \sum_{y>x} d^2 _{y,x}  \langle A^\perp \rangle \geq 2\frac{(\mu' _j)^2}{\langle \Sigma \rangle }.
\end{equation}
It is important to notice that  \eqref{eq: gen series activity bound} is valid for any graph, unlike \eqref{eq:two useful EP bounds} which are valid for the case of a $1D$ chain.

At this point, it is unclear  whether  $\mathcal{F}$ relates to either $\mu(\lambda)-\lambda \mu'(0)$, $\mu''(0)$.  In what follows, we discuss a special case  where one can order the bounds as already noticed in \cite{Barato_Gabrielli}. Furthermore, we show that $\mathcal{F}$ and $\mu''(0)$ do not bound one another.

\subsection{Application of TUR in 1D lattice -- A series of bounds \label{subsec:application 1d lattice}}
Consider an  1D lattice of $L$ sites which can be either open (boundary driven) or with periodic boundary conditions. Other boundary conditions can also be treated. We furthermore assume only nearest neighbors jumps \footnote{In fact, short range jumps is a sufficient conditions, but more tedious to treat and harder to precisely define on a short 1D chain. }. 
In this case, there is a single solution to Kirchhoff's junction rule \eqref{eq:Kirchhoffs rule}: $Y  \equiv Y_{x+1,x} $ for any $x$ on the lattice.  This in turn implies that the bound \eqref{eq:m2 activity bound} is indeed optimal in the $1D$ setup.
Notice that the summation depends on the boundary conditions in question. Since the activity bound is optimal (see the appendix \ref{app:derivation series of bounds} for a direct proof), we can also order the activity and entropy production bounds to follow \eqref{eq:series of bounds}.
Numerical evidence for the validity and relevance of these bounds were shown in Figs.~\ref{fig:mu A difference}~and \ref{fig: A sigma difference}.

The large deviation bound \eqref{eq: Y activity LDF bound} is not optimal even in the $1D$ case. Therefore, it is unclear whether one can order the large deviation bounds [\eqref{eq:LDF bound EP} and \eqref{eq: Y activity LDF bound}] in a similar fashion. Nevertheless, it is clear that in processes with even a single  unidirectional rate (no local equilibrium) $\langle  \Sigma \rangle \rightarrow \infty $ and  the bound \eqref{eq:LDF bound EP}  becomes irrelevant. In this case, clearly the large deviation bound \eqref{eq: Y activity LDF bound} dominates. Furthermore, in the $1D$ case $\langle j_{y,x}\rangle = J $ for any $y>x$. This further simplifies the evaluation of \eqref{eq: Y activity LDF bound}.

We stress again that \cite{Barato_Gabrielli} proved a similar bound to \eqref{eq:series of bounds}  even for finite $\mathcal{N}$. However, at large $\mathcal{N}$ values, the bound \eqref{eq:series of bounds} becomes tight and indeed a more informative bound to explore the entropy production. Moreover, here we consider the coarse grained densities instead of the densities that span over the full state space -- this renders a major advantage in the application of the bounds in many body systems.

Lastly, let us consider $\mathcal{F}=\sum_{x,y>x} d^2 _{y,x} \langle a_{y,x} \rangle$, which is a combination of the average activities. Since $\mathcal{F} \geq 2  \max_Y F_Y$ and $\mu''(0) \geq 2 \max_Y F_Y $ one may conjecture another bound $ \mathcal{F} \geq \mu''(0) $.
We have tested this conjecture numerically in Fig.~\ref{fig:curly F  mu }. For most  random realizations, the conjecture holds for any $\mathcal{N}$. However, a fraction of the realizations indeed exhibits violations of the conjecture which does not decrease with larger $\mathcal{N}$ values.


\section{Discussion and Summary}
\label{Sec:Discussion}


Inferring entropy production of an irreversible system is a central quest in biological systems and in thermal engines. Only recently, a major breakthrough has come through the field of stochastic thermodynamics --- a set of relations namely the TUR and subsequent results have been derived which show that the entropy production could be bounded by current fluctuations.
Nevertheless, current fluctuations are only easily accessible in an effective single body problem and specific solvable models. In many body systems, trading the difficulty of assessing the entropy production in assessing the current fluctuations usually means trading one difficult problem with another. Therefore, it is of interest to find meaningful accessible bounds to the entropy production in many-body systems. This work exactly addresses this question.

Our approach to study bounds on current fluctuations and the entropy production is based on a large $\mathcal{N}$ theory spanned over finite graphs.
We show that the current variance \eqref{eq:m2 activity bound} as well as the cumulant generating function \eqref{eq: Y activity LDF bound} can be bounded by the coarse grained activities that are given in terms of the steady state densities. Moreover, the entropy production is bounded by a the total activity in the system \eqref{eq: gen series activity bound}. Generally and on an arbitrary graph, the entropy production bounds  \eqref{eq: gen series activity bound} and the TUR \eqref{eq:LDF bound EP} cannot be ordered. Nevertheless, it is clear that if the number of sites on the graph becomes large, the bound \eqref{eq: gen series activity bound} becomes irrelevant. This is because the entropy production is proportional to the volume of the graph and it is bounded by a term scaling like the inverse of the volume. Namely, \eqref{eq: gen series activity bound} is particularly relevant in small graphs with large occupancy.

Additionally, in 1D systems, we have shown that a series of bounds exists for the current variance, the activities and the entropy production. Surprisingly, the entropy production of the entire system can be bounded from the information in a single bond \eqref{eq:two useful EP bounds}. To gain further insights on these results, we have further studied an interacting model system namely the Asymmetric Inclusion Process (ASIP) and demonstrated that the activity bound is a significantly better bound for the entropy production when a large $\mathcal{N}$ limit is taken.

The large $\mathcal{N}$ theory assumes fast transition rates and large occupancies. Our results are valid within this framework, with a controlled error scaling like $1/\mathcal{N}$. This implies that our results could give a feasible estimate also for finite $\mathcal{N} $ values. Moreover, other scaling approaches could be considered in the large $\mathcal{N}$ limit \cite{Baek_largeN}, probably resulting in different bounds; such a possibility is left for future studies. 

The series of bounds is an appealing result as it suggests that fluctuating quantities can have useful bounds both from above and below. It remains to be seen whether one can bound, e.g. the entropy production from above as well as from below.  Another interesting avenue is an inverse problem of constructing networks such that useful series of bounds are obtained, following \eqref{eq:nu linear set}. It is also of interest to explore what is the family of networks (besides the 1D case) where \eqref{eq:series of bounds} still applies. 

We note that our work could be extended beyond steady state physics into the realm of periodically driven systems in the large $\mathcal{N}$ limit similar to \cite{Barato_Gabrielli}. Moreover, we expect the bounds derived here to remain relevant also close to phase transitions as well as dynamical phase transitions \cite{Gabrielli_LargeN,Don_DPT,Geo_DPT,Nemoto_DPT}. This statement might be surprising since at this regime fluctuations dominate and one may expect that finite $\mathcal{N}$ corrections to $\Phi $ are important. While this is true, it was already established that universal theories capture the relevant corrections close to a phase transition, i.e. the universal scaling function is attained \cite{Universality_DPT18,gerschenfeld2011current}.  Nevertheless, it would be interesting to explore the saturation of the bound close to a phase transition.

Designing principles consistent with thermodynamics in interacting particle systems leading to phenomena such as organization and self-assembly is an important challenge \cite{Suri_review}. Dynamic instability of biological machines such as microtubles is another such example where microtubules can grow and shrink from a centrosome in different tracks following absorption and escape of tubulins \cite{howard2003dynamics,howard2009growth}. The lattice ASIP model that has been studied here is a crude and elementary version of microtubules where particles play the role of tubulins. In this paper, we have shown how the uncertainty relations derived herein can be useful to provide informative bounds for interacting systems with large occupancies. Future studies need to be made to see whether such statistical model systems can be useful inspirations to unravel thermodynamic complexities in biological machines.

\section{Acknowledgements}
AP is indebted to Tel Aviv University (via the Raymond and Beverly Sackler postdoc fellowship and the fellowship from Center for the Physics and Chemistry of Living Systems) where the project started.

\appendix 

\section{Formulation of the probability distribution   and \eqref{eq:2point LDF}}
\label{app:probability distribution}

The purpose of this section is to derive the probability distribution $P(j,\rho)$  [more specifically to obtain $\Phi(j,\kappa_+,\kappa_-)$ as in \eqref{eq:2point LDF}] in the large $\mathcal{N}$ limit. 
To this end, let us
consider a jump process of two sites $x,y$. The rate for a particle to jump from $x \rightarrow y $ is denoted $\tilde{\kappa}_+ $ and to jump from   $y \rightarrow x $ is denoted $\tilde{\kappa}_- $. We wish to find the cumulant generating function of 
$
 e^{\lambda Q_+ dt - \lambda Q_- dt}
$ where $Q_\pm$ counts the number of jumps from $ x \rightleftarrows
 y $. Now,
expanding the cumulant generating function leads to 
\begin{align}
 &e^{\lambda Q_+ dt - \lambda Q_- dt} 
 = \nonumber \\ 
&    \left[ e^\lambda \tilde{\kappa}_+ dt + (1-\tilde{\kappa}_+ dt)   \right] 
    \left[ e^{-\lambda} \tilde{\kappa}_- dt + (1-\tilde{\kappa}_- dt)   \right]. 
\end{align}
Using the large $\mathcal{N}$ scaling $\tilde{\kappa}_\pm  =  \mathcal{N}^2 \kappa_\pm (\rho) + O(\mathcal{N})$  and $dt = d\tau /\mathcal{N}$, we find  
\begin{equation}
    e^{\lambda Q_+ dt - \lambda Q_- dt} = e^{\mathcal{N}d\tau \mu(\lambda)}, 
\end{equation}
where 
\begin{align}
    \mu(\lambda) = (e^{\lambda}-1)\kappa_+  + (e^{-\lambda}-1)\kappa_-.
\end{align}
 The saddle point at large $\mathcal{N}$ insures that the probability distribution $P(j,\rho) \sim \exp{(- \mathcal{N} \int d\tau \mathcal{L})} $ is connected to the cumulant generating function by a Legendre transform $\mathcal{L} = \lambda j - \mu(\lambda)$. Through the Legendre transformation we find 
 \begin{align}
   \lambda (j) &= \log \frac{j+\sqrt{j^2+4\kappa_+ \kappa_-}}{2\kappa_+} ,\\
   \mathcal{L} &= \Phi(j,\kappa_+,\kappa_-),
 \end{align}
 where $\Phi$ is given by \eqref{eq:2point LDF}.
 Note that the generalization to multiple sites and larger $\tau$ times is straight-forward.

\section{Derivation of the series of bounds  }
\label{app:derivation series of bounds}
In the main text, we have shown that $\mu''(0)\geq (d^\perp)^2\langle A^\parallel \rangle $. Furthermore, it is by now well known that $\mu''(0) \geq 2 (\mu' _j)^2/ \langle \Sigma \rangle $ \cite{Barato_Seifert15,Gingrich_PRL16}. Thus, we are left to prove that indeed $(d^\perp)^2 \langle A^\parallel \rangle  \geq 2 (\mu' _j )^2 / \langle \Sigma \rangle $ for 1D systems. 

To prove the inequality, we use $J = \langle j_{x+1,x} \rangle$ for any $x$ in the 1D setup. First, define
$r_x \equiv \langle a_{x+1,x} \rangle / |J| \geq 1 $. Second, note that 
\begin{align}
 \frac{1}{2} r \log \frac{r+1}{r-1} \geq 1  ~~\text{for}~~~ r>1.   
\end{align}
Then, it simply follows
\begin{equation}
    \frac{\langle A^\parallel \rangle \Sigma }{(2\mu' _j )^2} = \frac{\sum_x \log(\frac{r_x +1}{r_x -1}) }{\sum_x \frac{2}{r_x}} \geq 1.  
\end{equation}
as claimed in \eqref{eq:series of bounds}.

\section{Details of the numerical exploration of the bounds as presented in the main text}
\label{app:Numerically exploring the bounds}

In this section, we provide further details of the numerical simulations that were conducted to demonstrate our bounds in the main text. We consider an ASIP model with $L=3$ sites and $N=2$ particles with periodic boundary conditions. At each realization, we randomize the asymmetry rates $p_\pm (x) \in \left[0,1\right] $ for $x=1,2,3$. We consider the total charge flux on the ring $Q = \sum_{x=1,2,3} j_{x+1,x}$. Namely, $d_{x+1,x} =1, u_x=0 $.

For the small $N=2,L=3$ values, it is straight-forward to write the $6\times 6$ Markov matrix $M$ as well as the tilted matrix allowing to calculate the cumulants \cite{DerridaDoucot}. 
In the tilted matrix $M$, we redefine the Markov coefficients $M_{x,y} \rightarrow e^{\lambda q} M_{x,y}  $ when the transition leads to the flux $q$. The largest eigenvalue of the tilted Markov matrix corresponds to the cumulant generating function $\mu(\lambda)$. From $\mu(\lambda)$, the first two cumulants are directly accessible by differentiation. So, the current and current variance values obtained in the numerical procedure are exact and not a large $\mathcal{N}$ approximation. One can also define a tilted matrix for the activities to obtain them exactly.   The densities and entropy production can be directly evaluated using the eigenstate corresponding to the zero eigenvalue of the Markov matrix. This eigenstate corresponds to the steady state occupancies. By following this procedure we would recover 
the series of inequalities as in \cite{Barato_Gabrielli}. 


However, we wish to show that \eqref{eq:series of bounds} exists at the large $\mathcal{N}$ limit with $1/\mathcal{N}$ corrections. 
For the activities and currents, we use the large $\mathcal{N}$ result 
\begin{eqnarray}
    \langle a_{x+1,x} \rangle &=& p_+(x) \langle \rho_x \rangle  (1+\langle \rho_{x+1}\rangle )
    \\ \nonumber 
    &+& p_-(x) \langle \rho_{x+1} \rangle  (1+\langle \rho_{x}\rangle ) 
    \\ \nonumber 
    \langle j_{x+1,x} \rangle &=& p_+(x) \langle \rho_x \rangle  (1+\langle \rho_{x+1}\rangle )
    \\ \nonumber 
    &-& p_-(x) \langle \rho_{x+1} \rangle  (1+\langle \rho_{x}\rangle ) 
    \end{eqnarray}
    The steady state densities $\langle \rho_x \rangle $ are evaluated from the eigenstate corresponding to the zero eigenvalue of the Markov matrix. Finally, the entropy production is evaluated by the steady state currents and steady state activities discussed above. 
This mean field approach is naturally valid in the large $\mathcal{N}$ limit. Therefore, violations of the bounds may be expected.    

We consider in the Figs. (\ref{fig:mu A difference}) and (\ref{fig: A sigma difference}), the ASIP process with $200 $ realizations of the random rates. It becomes apparent that the two bounds become tight at the large $\mathcal{N}$ limit. Furthermore, we showed that indeed the bounds become tight with $1/\mathcal{N}$ corrections. This can be validated by considering the minimal differences. The plots indeed show that
\begin{align}
    \frac{\mu''(0)-(d^\perp) ^2 \langle A^\parallel \rangle }{\mu''(0)} ,\frac{(d^\perp) ^2 \langle A^\parallel \rangle - 2 (\mu' _j )^2 /\Sigma   }{(d^\perp) ^2 \langle A^\parallel \rangle} \sim  1/\mathcal{N} ,
    \label{minimal-differences}
    \end{align}
as predicted theoretically. We note that while the variance-activity bound can be both positive or negative for finite $\mathcal{N}$, the activity-entropy production bound is strictly non-negative for any $\mathcal{N}$. This need not persist for any dynamics. However, note that for each local activity $\langle \frac{1}{a_{x,y}} \rangle \geq \frac{1}{\langle a_{y,x} \rangle }$, making violations uncommon in finite $\mathcal{N}$. Equality is reached only in the large $\mathcal{N}$ limit. Together with the  \cite{Barato_Gabrielli} bound,  the positivity of the bound for any $\mathcal{N}$ is therefore not surprising.


\section{Numerical examination of the bound $\mathcal{F} \geq \mu''(0)$ 
\label{app: the F numerics} 
}  

\begin{figure}
     \centering
     \begin{subfigure}[b]{0.3\textwidth}
         \centering
         \includegraphics[width=\textwidth]{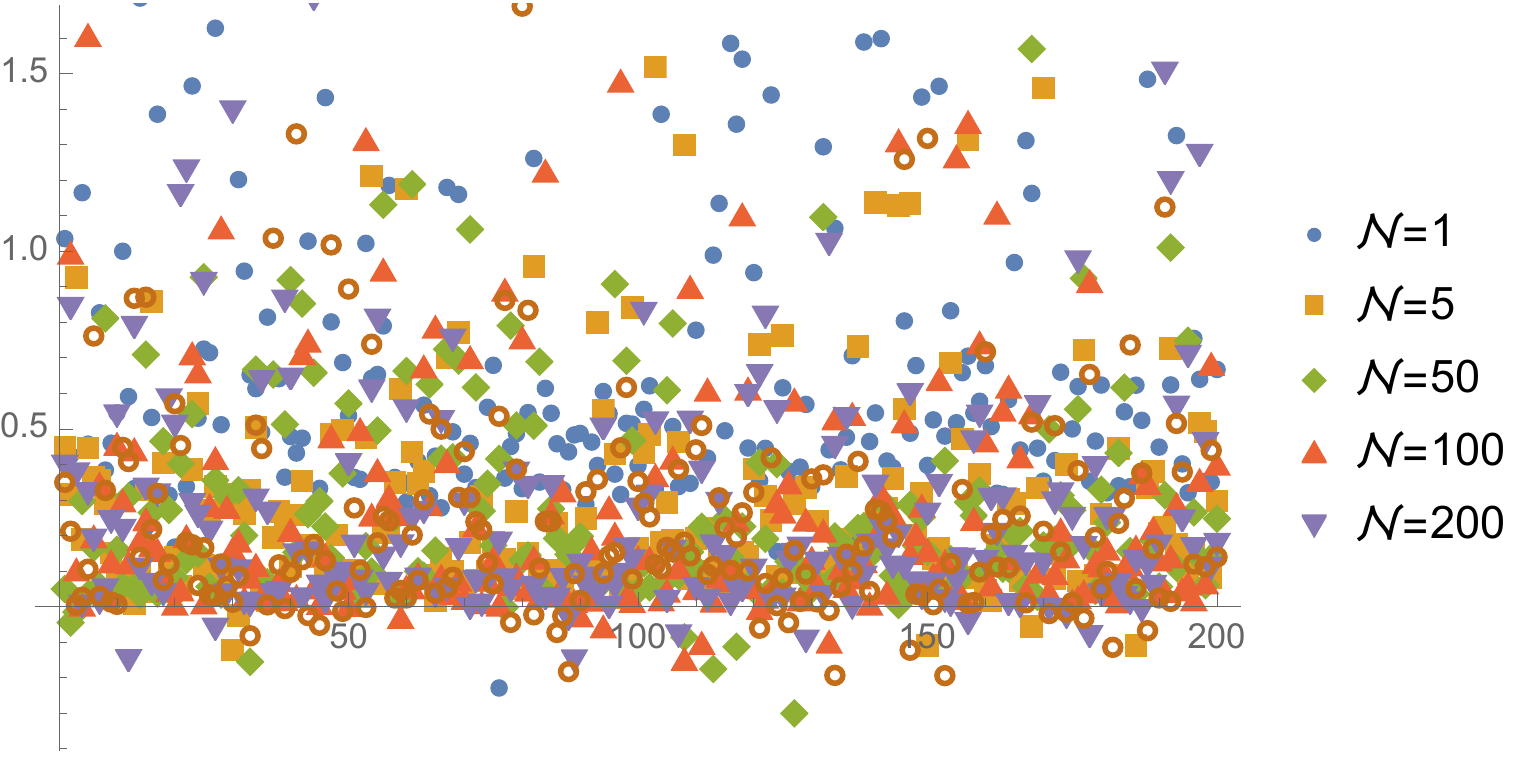}
         \caption{200 random realizations of the ASIP }
     \end{subfigure}
     \hfill
     \begin{subfigure}[b]{0.3\textwidth}
         \centering
     \includegraphics[width=\textwidth]{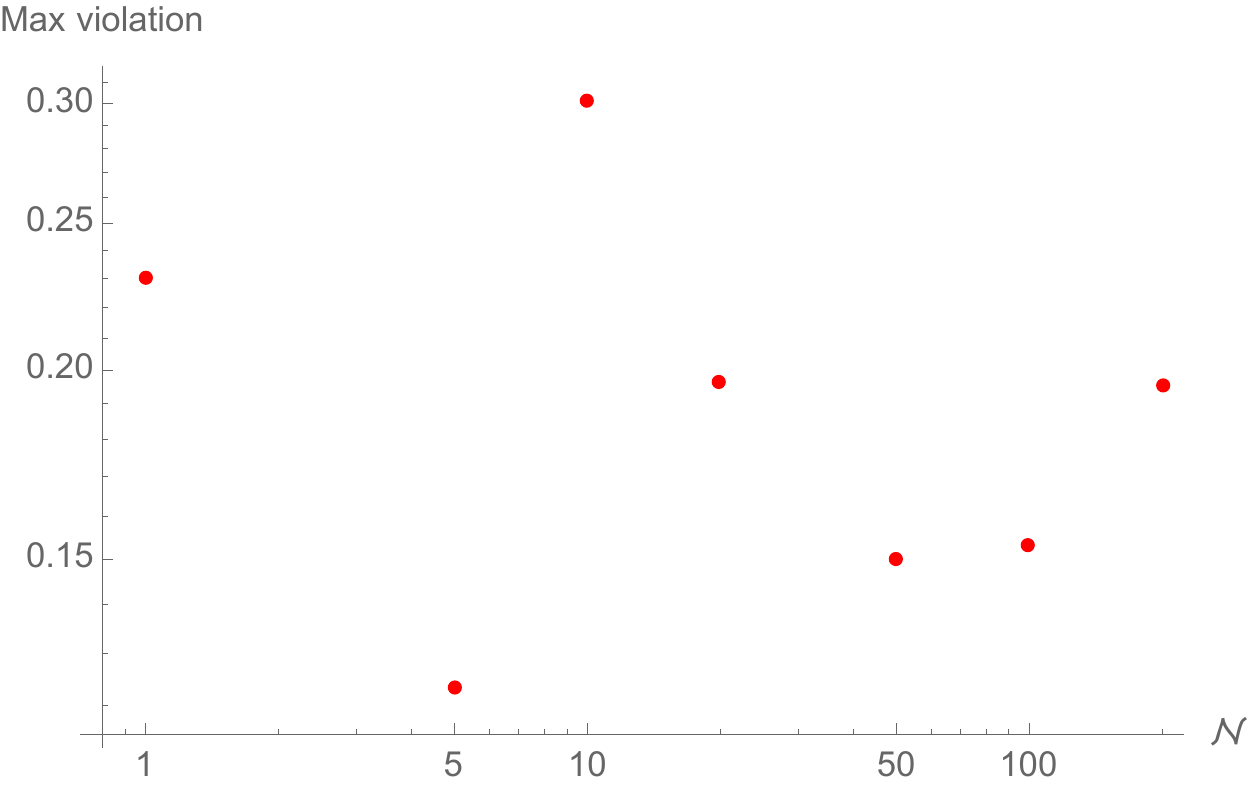}
         \caption{ The negative of minimal value of $\frac{\mathcal{F}-\mu''(0)}{\mu''(0)}$  out of all the random realizations for each $\mathcal{N}$ -- the max violation.    
         }
     \end{subfigure}
     \hfill
        \caption{The value of $\frac{\mathcal{F}-\mu''(0)}{\mu''(0)}$ in the random ASIP of $N=2$ particles and $L=3$ sites with periodic boundary conditions. We present $200$ random realizations for each $\mathcal{N}$ value, where the rates $p_\pm(x)\in \left[ 0,1\right]$ are randomized.   
        (a) While most realizations at any $\mathcal{N}$ satisfy the bound, there are violations. The maximal violations of each $
        \mathcal{N}$ are plotted in (b). They does not vanish with increasing $\mathcal{N}$.  }
        \label{fig:curly F  mu }
\end{figure}

In the main text (Sec. \ref{subsec:application 1d lattice}) it was argued that $\mathcal{F} \geq \mu''(0) $ is satisfied for most realizations, but not all. Furthermore, this statement does not depend on the $\mathcal{N}$ value. Here we present numerical evidence for this claim. See Fig.~\ref{fig:curly F  mu } and the captions therein.

\section{Derivation of the series activity bound: the kinetic uncertainty relation}
\label{app:series activity bound}
In this section, we use \eqref{eq:nonphysical bound} derive a bound on the current variance in terms of the series activity $\langle A^\perp \rangle $ also known in the literature as the kinetic uncertainty relation \cite{KUR_Baiesi,Shiraishi_Optimalbound,UnidrectionalTUR}. Let us start from \eqref{eq:nonphysical bound} and take $Y_{x,y} = \alpha \langle j_{x,y} \rangle$ for some constant $\alpha$. Notice that this choice immediately satisfies the Kirchhoff's junction rule \eqref{eq:Kirchhoffs rule}. From \eqref{eq:nonphysical bound}, we then find the following inequality
\begin{align}
    \frac{1}{2}\mu''(0) &\geq \alpha \mu' _j - \alpha^2 \sum_{x,y>x} \frac{\langle j_{y,x} \rangle^2 }{\langle a_{x,y} \rangle }
    \\ \nonumber 
    & = \alpha \mu' _j - \alpha^2 \sum_{x,y>x} \langle a_{x,y} \rangle \frac{\langle j_{y,x} \rangle^2 }{\langle a_{x,y} \rangle^2 }. 
\end{align}
Notice that since $|\langle j_{y,x} \rangle / \langle a_{x,y} \rangle | \leq 1$, we obtain 
\begin{equation}
    \frac{1}{2} \mu''(0) \geq \alpha \mu' _j - \frac{1}{2}\alpha^2  \langle A^\perp \rangle. 
\end{equation}
Now, it is straight-forward to choose $\alpha = \mu' _j / \langle A^\perp \rangle$ to obtain the kinetic uncertainty bound 
\begin{equation}
    \mu''(0) \geq \frac{(\mu' _j)^2 }{ \langle A^\perp \rangle} ,
\end{equation}
that was obtained in \cite{KUR_Baiesi,Shiraishi_Optimalbound,UnidrectionalTUR}.

\bibliography{refs} 

\end{document}